\documentclass[10pt,aps,onecolumn,nofootinbib]{revtex4}

\usepackage[bigfiles]{pdfbase}
\usepackage[normalem]{ulem}
\ExplSyntaxOn
\NewDocumentCommand\embedvideo{smm}{
  \group_begin:
  \leavevmode
  \tl_if_exist:cTF{file_\file_mdfive_hash:n{#3}}{
    \tl_set_eq:Nc\video{file_\file_mdfive_hash:n{#3}}
  }{
    \IfFileExists{#3}{}{\GenericError{}{File~`#3'~not~found}{}{}}
    \pbs_pdfobj:nnn{}{fstream}{{}{#3}}
    \pbs_pdfobj:nnn{}{dict}{
      /Type/Filespec/F~(#3)/UF~(#3)
      /EF~<</F~\pbs_pdflastobj:>>
    }
    \tl_set:Nx\video{\pbs_pdflastobj:}
    \tl_gset_eq:cN{file_\file_mdfive_hash:n{#3}}\video
  }
  \pbs_pdfobj:nnn{}{dict}{
    /Type/RichMediaInstance/Subtype/Video
    /Asset~\video
    /Params~<</FlashVars (
      source=#3&
      skin=SkinOverAllNoFullNoCaption.swf&
      skinAutoHide=true&
      skinBackgroundColor=0x5F5F5F&
      skinBackgroundAlpha=0.75
      loop=True
    )>>
  }
  \pbs_pdfobj:nnn{}{dict}{
    /Type/RichMediaConfiguration/Subtype/Video
    /Instances~[\pbs_pdflastobj:]
  }
  \pbs_pdfobj:nnn{}{dict}{
    /Type/RichMediaContent
    /Assets~<<
      /Names~[(#3)~\video]
    >>
    /Configurations~[\pbs_pdflastobj:]
  }
  \tl_set:Nx\rmcontent{\pbs_pdflastobj:}
  \pbs_pdfobj:nnn{}{dict}{
    /Activation~<<
      /Condition/\IfBooleanTF{#1}{PV}{XA}
      /Presentation~<</Style/Embedded>>
    >>
    /Deactivation~<</Condition/PI>>
  }
  \hbox_set:Nn\l_tmpa_box{#2}
  \tl_set:Nx\l_box_wd_tl{\dim_use:N\box_wd:N\l_tmpa_box}
  \tl_set:Nx\l_box_ht_tl{\dim_use:N\box_ht:N\l_tmpa_box}
  \tl_set:Nx\l_box_dp_tl{\dim_use:N\box_dp:N\l_tmpa_box}
  \pbs_pdfxform:nnnnn{1}{1}{}{}{\l_tmpa_box}
  \pbs_pdfannot:nnnn{\l_box_wd_tl}{\l_box_ht_tl}{\l_box_dp_tl}{
    /Subtype/RichMedia
    /BS~<</W~0/S/S>>
    /Contents~(embedded~video~file:#3)
    /NM~(rma:#3)
    /AP~<</N~\pbs_pdflastxform:>>
    /RichMediaSettings~\pbs_pdflastobj:
    /RichMediaContent~\rmcontent
  }
  \phantom{#2}
  \group_end:
}
\ExplSyntaxOff

\usepackage{epsfig}
\usepackage{amsmath,amsfonts,amssymb}
\usepackage{t1enc}
\usepackage{verbatim}
\usepackage{float}
\usepackage{morefloats}
\usepackage{color}
\usepackage{physics}
\usepackage{booktabs}
\usepackage[Q=yes,pverb-linebreak=no]{examplep}
\newcommand{\red}[1]{\textcolor{red}{#1}}
\newcommand{\blue}[1]{\textcolor{blue}{#1}}
\begin{document}

\title{\boldmath The Dirac-Delta Rogue Wave \unboldmath}

\author{
Miguel C. N. Fiolhais$^{1,2,3}$ and Andrea Ferroglia$^{3,4}$
\\[3mm]
{\footnotesize {\it 
$^1$ Science Department, Borough of Manhattan Community College, \\ The City University of New York, 199 Chambers St, New York, NY 10007, USA \\ 
$^2$ LIP, Departamento de F\'{\i}sica, Universidade de Coimbra, 3004-516 Coimbra, Portugal \\
$^3$ The Graduate School and University Center, The City University of New York, 365 Fifth Avenue, New York, NY 10016  USA \\
$^4$ Physics Department, New York City College of Technology, The City University of New York, 300 Jay Street, Brooklyn, NY 11201, USA \\
}}
}

\begin{abstract}
This paper presents a double spatio-temporal localized Dirac-delta solution for the linear wave equation. 
The solution arises from the interference of sinusoidal waves with frequencies that vary as a function of the time of emission. It is shown that the time-dependent frequency function required to produce a localized Dirac-delta wave is exclusively determined by the dispersion relation of the medium in which the wave is traveling. Both numerical and exact analytical solutions for a typical physical scenario are obtained and displayed. The application of this result to possible practical uses is discussed.
\end{abstract}

\maketitle

\section{Introduction}
\label{sec:intro}

A rogue wave, also known as a freak wave or monster wave, is an exceptionally large ocean wave that occurs unexpectedly amid relatively smaller waves. Its immense size, steepness, and unpredictability make it extremely dangerous for ships, offshore structures and coastal areas. Rogue waves appear suddenly and can reach heights of more than twice the significant wave height (the average height of the highest third of waves in a given area.) Consequently, these waves can have devastating impacts on vessels, capable of capsizing or severely damaging even large ships~\cite{PhysRevA.80.043818,PhysRevLett.114.213901}.

Once considered mythical or exaggerated accounts by sailors, the existence of these waves has been confirmed through scientific observations and satellite data~\cite{christou,benetazzo,onoratoObservation}. One possible explanation stems from a solution of the non-linear  Schr\"{o}dinger equation - the Peregrine soliton~\cite{Peregrine1983}. {The non-linear  Schr\"{o}dinger equation, is a fundamental equation in quantum mechanics and nonlinear optics~\cite{kaertner}. It extends the Schrödinger equation to describe the behavior of complex wave phenomena in nonlinear systems and plays a crucial role in understanding the dynamics of Bose-Einstein condensates, optical solitons, and other nonlinear wave phenomena~\cite{bronski,ahmad}. In the context of water waves, it accounts for phenomena such as wave steepening, wave modulation, and wave interactions, allowing for a more accurate representation of the complex dynamics of waves in various conditions, such as those encountered in the ocean~\cite{onorato}.} In contrast to a typical soliton, which can retain its profile without alteration while propagating, the Peregrine soliton exhibits a unique double spatio-temporal localization. Consequently, when initiated with a modest oscillation against a continuous background, the Peregrine soliton experiences a gradual amplification of its intensity. Following this stage of maximum compression, the soliton's amplitude diminishes and its width expands. In other words, large amplitudes that seemingly materialize out of thin air, vanish without a trace just like a rogue wave. This phenomenon has been experimentally reproduced in water tanks experiments~\cite{PhysRevLett.106.204502,PhysRevX.2.011015} and has also been observed in a variety of physical systems~\cite{ONORATO201347}. In the linear regime however, for example in the approximate Airy wave theory for sea waves or in any other physical system that follows a linear wave equation, rogue waves may result purely from the constructive interference of smaller waves in dispersive media~\cite{2023PNAS..12006275H}. In certain conditions, waves traveling at different speeds may align and combine, leading to the formation of a larger, more powerful wave for a short period of time before disappearing onto the background. 

The objective of this paper is to explore the possibility of generating such effect with the emission of a sinusoidal wave with a varying frequency, such as a chirp, in a dispersive medium. In order to do so, a general solution of the classical linear wave equation is obtained for the extreme case of a wavepacket that converges into a Dirac-delta function at a fixed location for an arbitrary dispersive medium. The analytical solution is obtained by considering the superposition of an infinite succession of infinitesimal
wavefronts generated at the source with a changing frequency. The same exact solution is also derived by using the short-time Fourier transform of the wave at the source, and by propagating each sinusoidal component of the transform as a plane wave according to the dispersion relation of the medium. In addition, it is shown that the time-dependent frequency function is determined by the dispersion relation, making the production of a Dirac-delta rogue wave possible in any dispersive medium, as long as the varying frequency of the emitted sinusoidal wave is adjusted accordingly. 

The results for the analytical solution are confirmed numerically for the particular case of a dispersion relation where the angular frequency is proportional to the square of the wavenumber. The numerical approach is employed by considering an emission signal at the source comprising a continuous succession of very short but finite sinusoidal signals, in order to simulate a chirp. These plane waves are propagated into the medium according to the dispersion relation, and their superposition is directly compared with the analytical solution. The relevance and possible application of the result presented in this paper is discussed in different contexts.



\section{Localized Dirac-Delta Wave}
\label{sec:II}


Consider a plane sinusoidal wave that is emitted along an $x$-axis with a time-varying frequency. For simplicity, the source of the wave is made to coincide with the origin of the $x$-axis. In addition, the amplitude at the source at the instant  $t_s$ is given the following value:
\begin{equation}
F(x=0,t_s) = \sin \varphi(t_s) \, ,
\label{eq:sourcewave}
\end{equation}
where the instantaneous angular frequency at the source is given by the time derivative of the angular phase,
\begin{equation}
\omega(t_s) = \frac{\partial\varphi(t_s)}{\partial t_s} \, .
\end{equation} 

\begin{figure}[h]
    \centering
\includegraphics[width=0.49\textwidth]{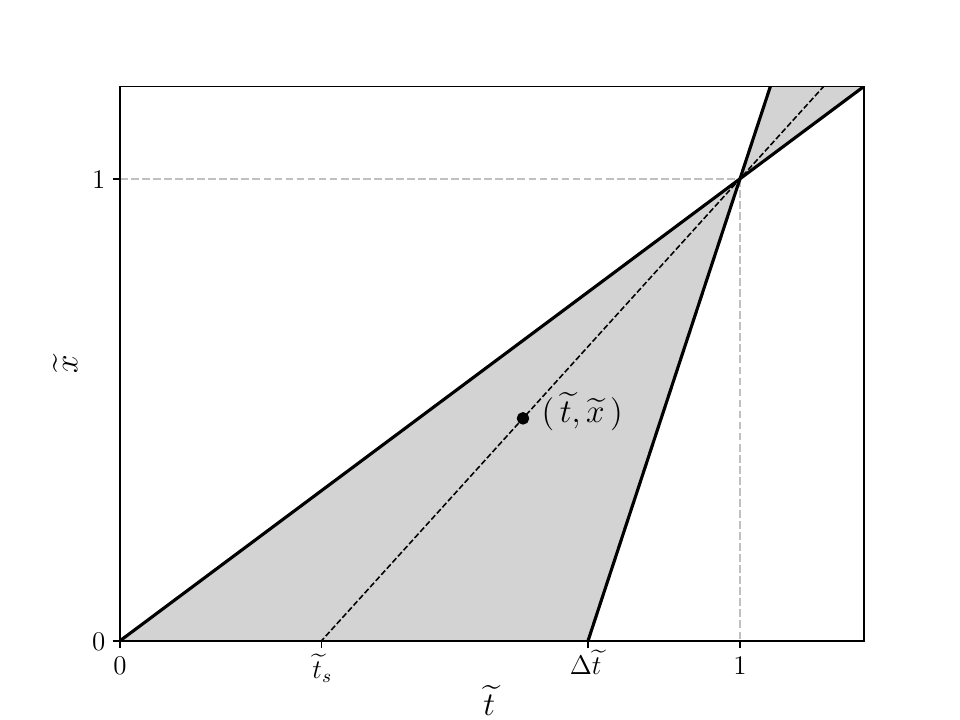}
\includegraphics[width=0.49\textwidth]{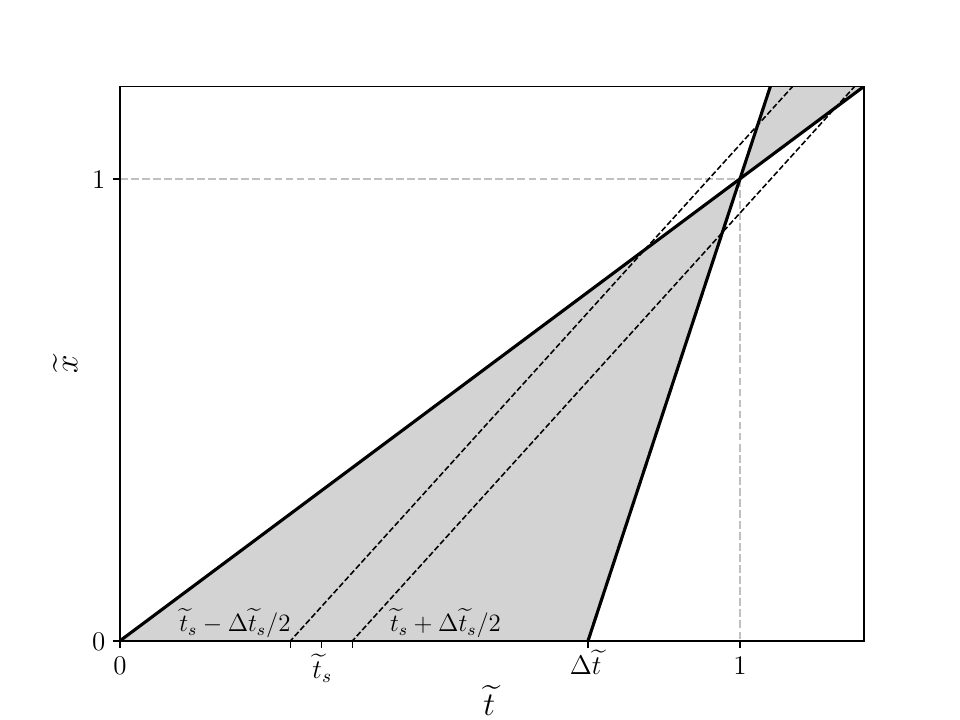}
    \caption{Dimensionless position-time diagrams for the emission of a wave during the interval $\Delta\widetilde{t}$ and converging at $\widetilde{x} =\widetilde{t} =1$. The left panel shows the path of a phase emitted from the source at the instant $\widetilde{t_s}$ as a dashed line, while the right panel shows the same phase emitted with a fixed frequency in the finite interval between $\widetilde{t}_s - \Delta \widetilde{t}_s / 2$ and $\widetilde{t}_s + \Delta \widetilde{t}_s / 2$.}
    \label{fig:diagram}
\end{figure}

In a dispersive medium, if the frequency at the source is varied over time, successive fronts of the wave will be transmitted with a changing instantaneous frequency at different phase velocities. If the variation of the frequency is continuous, each wavefront emitted through an infinitesimal
time interval can be made to arrive simultaneously at a point $x=L$ at a time instant $t_0$. 
By considering the wavefront emitted from the source at the time instant $t_s$, the required velocity for it to arrive at $x=L$ can be determined as,
\begin{equation}
v = \frac{\omega}{k} = \frac{L}{t_0-t_s} \, .
\label{eq:velocity1}
\end{equation}
The convergence of phases is perhaps better observed in the diagrams shown in Fig.~\ref{fig:diagram}, where the position and time coordinates are represented in the vertical and horizontal axes, respectively. For convenience, dimensionless variables are used for position and time, defined as:
\begin{equation}
\widetilde{x} = \frac{x}{L} \,\,\,\,\,\,\,\,\,\,\,\,\,\,\,\,\,\,\,\,\, \textrm{and}  \,\,\,\,\,\,\,\,\,\,\,\,\,\,\,\,\,\,\,\,\, \widetilde{t} = \frac{t}{t_0}  .
\end{equation}
\newline
If the dispersion relation of the medium, $\omega(k)$, is known, then Eq.~(\ref{eq:velocity1}) fully determines the time dependency of frequency at the source $\omega(\widetilde{t}_s)$. Note that, for convenience, Eq.~(\ref{eq:velocity1}) can also be expressed in terms of a dimensionless angular frequency, $\widetilde{\omega}=\omega/\omega_0$, and wavenumber, $\widetilde{k}=k/k_0$,
\begin{equation}
\frac{\widetilde{\omega}}{\widetilde{k}} = \frac{k_0 L}{\omega_0 t_0} \frac{1}{1-\tilde{t}_s} \, ,
\label{eq:velocity2}
\end{equation}
where $k_0$ is the wavenumber corresponding to the angular frequency $\omega_0$ in a given dispersive medium. For simplicity, the special case where the values of $\omega_0 t_0$ and $k_0 L$ are equal to one will be considered for the remainder of this section.

As the time-dependent frequency at the origin is found directly from the dispersion relation, the next step is to calculate the overall wave as a result of the superposition of the infinite number of wavefronts emitted during infinitesimal time intervals $\textrm{d}t_s$ between the initial time instant and an arbitrary time interval $\Delta t$, with $\Delta t < t_0$. 

The equation of a wavefront emitted from the origin at a time $\widetilde{t}_s$ can be written as,
\begin{equation}
f(\, \widetilde{x},\widetilde{t}\, ) = \mathcal{U}\left( \, \widetilde{t}_s - \frac{\widetilde{t}-\widetilde{x} }{1-\widetilde{x}} \, \right) \sin\left[ \widetilde{\omega}(\widetilde{t}_s) \widetilde{t} - \widetilde{k}(\widetilde{t}_s)  \widetilde{x} + \phi(\widetilde{t}_s) \right] \, ,
\end{equation}
for an arbitrary point $(\widetilde{t},\widetilde{x})$ in the left diagram of Fig.~\ref{fig:diagram}, where
\begin{equation}
\mathcal{U}(z)=
    \begin{cases}
        1 & \text{if } z =0 \, , \\
        0 & \text{if }  z\neq 0 \, .
    \end{cases}
\end{equation}
Note that the relation,
\begin{equation}
\widetilde{t}_s= \frac{\widetilde{t}-\widetilde{x} }{1-\widetilde{x}} \, ,
\end{equation}
corresponds to the equation of the dashed line in the left graph of Fig.~\ref{fig:diagram}, where $f$ takes a constant value along the line and is zero elsewhere. In addition, as there is no superposition at the source, the phase $\phi(\widetilde{t}_s)$ is determined by the boundary condition,
\begin{eqnarray}
f(\, 0,\widetilde{t}_s\, ) & = & F(\, 0,\widetilde{t}_s\, ) \\
\sin\left[ \widetilde{\omega}(\widetilde{t}_s) \widetilde{t}_s + \phi(\widetilde{t}_s) \right] & = & \sin \varphi(\widetilde{t}_s) \, ,
\end{eqnarray}
which leads to,
\begin{equation}
\phi(\widetilde{t}_s) =  \varphi(\widetilde{t}_s) - \widetilde{\omega}(\widetilde{t}_s) \widetilde{t}_s \, .
\label{eq:phiphase}
\end{equation}

According to the superposition principle, the overall wave can then be determined as the infinite sum of individual wavefronts emitted from the source over the interval $\Delta \widetilde{t}$. To this effect, it is convenient to perform this sum over small but finite intervals $\Delta \widetilde{t}_s$, as represented on the right diagram of Fig.~\ref{fig:diagram}, 
and then take the limit when these intervals become infinitesimal ($\Delta \widetilde{t}_s \to \textrm{d} \widetilde{t}_s$), so that the interference is described by an integral. For instance, if a wave is being emitted with a fixed angular frequency ${\omega}(\widetilde{t}_s)$ in the interval between $\widetilde{t}_s - \Delta \widetilde{t}_s / 2$ and $\widetilde{t}_s + \Delta \widetilde{t}_s / 2$, the function $\mathcal{U}$ can be expressed as,
\begin{equation}
\mathcal{U}\left( \, \widetilde{t}_s - \frac{\widetilde{t}-\widetilde{x} }{1-\widetilde{x}} \, \right)=
    \begin{cases}
        1 & \text{if }  -\frac{\Delta \widetilde{t}_s }{ 2|1-\widetilde{x}|} <\widetilde{t}_s - \frac{\widetilde{t}-\widetilde{x} }{1-\widetilde{x}} < +\frac{\Delta \widetilde{t}_s }{ 2|1-\widetilde{x}|} \, , \\
        0 & \text{elsewhere,}
    \end{cases}
    \label{eq:U2}
\end{equation}
where the condition expressed by the inequality in the equation above corresponds to the region between the two dashed lines on the right panel of Fig.~\ref{fig:diagram}. In the limit $\Delta \widetilde{t}_s \rightarrow 0$, these two parallel lines merge and the function $\mathcal{U}$ can be written in terms of the Dirac delta function as:
\begin{equation}
\lim_{\Delta\widetilde{t}_s \to 0} \frac{1}{\Delta\widetilde{t}_s} \, \mathcal{U}\left( \, \widetilde{t}_s - \frac{\widetilde{t}-\widetilde{x} }{1-\widetilde{x}} \, \right) = \frac{1}{|1-\widetilde{x}|} \delta\left( \, \widetilde{t}_s - \frac{\widetilde{t}-\widetilde{x} }{1-\widetilde{x}} \, \right)  \, .
\end{equation}
As a result, the final superposed wave in the continuous limit is,
\begin{equation}
F(\, \widetilde{x},\widetilde{t}\, ) = \frac{1}{|1-\widetilde{x}|} \int_0^{\Delta \widetilde{t}} \mathcal{\delta}\left( \, \widetilde{t}_s - \frac{\widetilde{t}-\widetilde{x}}{1-\widetilde{x}} \, \right) \sin\left[ \widetilde{\omega}(\widetilde{t}_s) \widetilde{t} - \widetilde{k}(\widetilde{t}_s)  \widetilde{x} + \phi(\widetilde{t}_s) \right] \textrm{d}\widetilde{t}_s \, .
\label{eq:integral}
\end{equation}
which in turn leads to the exact solution in a closed form:
\begin{equation}
F(\, \widetilde{x},\widetilde{t}\, ) = \frac{1}{|1-\widetilde{x}|} \sin\left[ \widetilde{\omega}\left(\frac{\widetilde{t}-\widetilde{x}}{1-\widetilde{x}}\right) \widetilde{t} - \widetilde{k}\left(\frac{\widetilde{t}-\widetilde{x}}{1-\widetilde{x}}\right)  \widetilde{x} + \phi\left(\frac{\widetilde{t}-\widetilde{x}}{1-\widetilde{x}}\right) \right] \, .
\label{eq:final1}
\end{equation}

By using Eqs.~(\ref{eq:velocity2}) and (\ref{eq:phiphase}), the superposition of waves in the continuous limit, expressed in Eq.~(\ref{eq:integral}), can also be rewritten as,
\begin{equation}
F(\, \widetilde{x},\widetilde{t}\, ) = \frac{1}{|1-\widetilde{x}|} \int_0^{\Delta \widetilde{t}} \mathcal{\delta}\left( \, \widetilde{t}_s - \frac{\widetilde{t}-\widetilde{x}}{1-\widetilde{x}} \, \right) \sin\left[ \widetilde{\omega}(\widetilde{t}_s) \left( \widetilde{t} - \widetilde{x} - (1-\widetilde{x}) \widetilde{t}_s \right) + \varphi(\widetilde{t}_s) \right] \textrm{d}\widetilde{t}_s \, ,
\label{eq:integralsub}
\end{equation}
which upon integration leads to a more compact form,
\begin{equation}
F(\,\widetilde{x},\widetilde{t}\,)=\frac{1}{|1-\widetilde{x}|} \sin  {\varphi} \left( \, { \frac{\widetilde{t}-\widetilde{x}}{1-\widetilde{x}}} \, \right)  \, .
\label{eq:final2}
\end{equation}

It is clear from this expression that the amplitude diverges at $\widetilde{x}=1$. It is important to stress that due to the superposition principle, the integral of the function over time at any fixed location must remain constant, \emph{i.e.} independent of the position - every wavefront emitted from the source over an infinitesimal time, must eventually cross any position in front of the source. Therefore, as the wave is radiated from the source, it converges into a Dirac delta function at a fixed location before dispersing into infinity - a \textbf{Dirac-delta rogue wave}. Naturally, if the integral is negative, the delta function is also negative at the divergence point, leading to a narrow deep depression instead of a towering wave.

Indeed, in the Appendix, it is proven 
that $F$ behaves as a Dirac-Delta function in both $\widetilde{x}$ and $\widetilde{t}$, so that for a smooth test function $h$ one finds that
\begin{align}
\lim_{\widetilde{t} \to 1} \frac{1}{G_x( \,\Delta \widetilde{t} \,)} & \int_{\frac{\widetilde{t} - \Delta \widetilde{t}}{1 -\Delta \widetilde{t}}}^{\widetilde{t}} \textrm{d} \widetilde{x} \,\,\, h(\,\widetilde{x}\,) F(\,\widetilde{x},\widetilde{t}\,) =  \int_{-\infty}^{\infty} \textrm{d} \widetilde{x} \,\,\, h(\,\widetilde{x}\,) \delta(\,1-\widetilde{x}\,)= h(1) \, , \nonumber \\
\lim_{\widetilde{x} \to 1} \frac{1}{G_t( \,\Delta \widetilde{t} \,)} & \int_{\widetilde{x}}^{\widetilde{x} + \Delta \widetilde{t} (1- \widetilde{x})}\!\!\!\! \textrm{d} \widetilde{t} \,\,\, h(\,\widetilde{t}\,) F(\,\widetilde{x},\widetilde{t}\,) =  \int_{-\infty}^{\infty} \textrm{d} \widetilde{t} \,\,\, h(\,\widetilde{t}\,) \delta(\,1-\widetilde{t}\,)= h(1) \, ,
\end{align}
where $G_x$ and $G_t$ are suitable normalization factors whose analytic expression is given in the Appendix.

\section{Short-Time Fourier Transform \label{sec:III}}

The same analytical solution presented in Section~\ref{sec:II} can also be derived using a short-time Fourier transform (STFT)~\cite{sejdic}. In order to do so, the signal at the source can be divided into shorter segments, and the Fourier transform can be calculated separately on each segment. As such, in the limit in which the short time intervals become infinitesimal, the signal at the source at a given time $\tau$ can be expressed as,
\begin{equation}
F(x=0,\tau) = \sin \varphi(\tau) = \int_0^{\Delta \widetilde{t}} \delta (\tau-\widetilde{t}_s) \sin \varphi(\widetilde{t}_s) \, \textrm{d}\widetilde{t}_s \, ,
\label{eq:stft0}
\end{equation}
where the Dirac delta $\delta (\tau-\widetilde{t}_s)$ can be interpreted as the window function centered around $\widetilde{t}_s$. The Fourier transform of the integrand function,
\begin{equation}
f(\, \tau,\widetilde{t}_s\, ) = \delta(\tau-\widetilde{t}_s) \sin  \varphi(\widetilde{t}_s)  \, ,
\label{eq:stft1}
\end{equation}
with respect to $\tau$ is therefore,
\begin{equation}
\mathcal{F}(\widetilde{\omega},\widetilde{t}_s)= \frac{1}{\sqrt{2\pi}}\int_0^{\Delta \widetilde{t}} \delta(\tau'-\widetilde{t}_s) \sin \varphi(\widetilde{t}_s) e^{-i \widetilde{\omega} \tau'} \,  \textrm{d}\tau' = \frac{1}{\sqrt{2\pi}} \sin\varphi(\widetilde{t}_s) e^{-i \widetilde{\omega} \widetilde{t}_s} \, ,
\label{eq:stft4}
\end{equation}
and the function can be recovered as,
\begin{equation}
f(\, \tau,\widetilde{t}_s\, ) = \frac{1}{\sqrt{2\pi}} \int_{-\infty}^{+\infty} \mathcal{F}(\widetilde{\omega},\widetilde{t}_s) e^{i \widetilde{\omega} \tau} \textrm{d}\widetilde{\omega} \, .
\label{eq:stft5}
\end{equation}
As each individual sinusoidal signal at a time $\tau$ at the source, that is multiplied to the Fourier transform, must propagate as a plane wave according to the dispersion relation the medium, the integrand function can be written in terms of $\widetilde{x}$ and $\widetilde{t}$ as,
\begin{equation}
f(\widetilde{x},\widetilde{t},\widetilde{t}_s) = \frac{1}{\sqrt{2\pi}} \int_{-\infty}^{+\infty} \mathcal{F}(\widetilde{\omega},\widetilde{t}_s) e^{i \widetilde{\omega} \widetilde{t} - i \widetilde{k} \widetilde{x}} \textrm{d}\widetilde{\omega} \, ,
\label{eq:stft6}
\end{equation}
where $\widetilde{k}$ is the wave number associated to the frequency $\widetilde{\omega}$. Consequently, the overall wave can also be calculated for any pair of $\widetilde{x}$ and $\widetilde{t}$ as,
\begin{equation}
F(\,\widetilde{x},\widetilde{t}\,) =  \int_0^{\Delta \widetilde{t}} f( \widetilde{x},\widetilde{t},\widetilde{t}_s) \, \textrm{d}\widetilde{t}_s = \frac{1}{{2\pi}} \int_0^{\Delta \widetilde{t}} \int_{-\infty}^{+\infty} \sin \varphi(\widetilde{t}_s) e^{-i \widetilde{\omega} \widetilde{t}_s + i \widetilde{\omega} \widetilde{t} - i \widetilde{k} \widetilde{x}} \, \textrm{d}\widetilde{\omega} \, \textrm{d}\widetilde{t}_s
\label{eq:stft6}
\end{equation}
This double integral can be rearranged as,
\begin{equation}
F(\,\widetilde{x},\widetilde{t}\,) = \int_0^{\Delta \widetilde{t}} \left ( \frac{1}{2\pi} \int_{-\infty}^{+\infty}  e^{-i \widetilde{\omega} \widetilde{t}_s + i \widetilde{\omega} \widetilde{t} - i \widetilde{k} \widetilde{x} } \, \textrm{d}\widetilde{\omega}  \right )  \sin \varphi(\widetilde{t}_s)  \,   \textrm{d}\widetilde{t}_s \, .
\label{eq:stft7}
\end{equation}
By imposing the condition of Eq.~(\ref{eq:velocity2}), the final wave becomes,
\begin{equation}
F(\, \widetilde{x},\widetilde{t}\, ) = \frac{1}{|1-\widetilde{x}|} \int_0^{\Delta \widetilde{t}} \mathcal{\delta}\left( \, \widetilde{t}_s - \frac{\widetilde{t}-\widetilde{x}}{1-\widetilde{x}} \, \right) \sin \varphi(\widetilde{t}_s)  \textrm{d}\widetilde{t}_s =  \, \frac{1}{|1-\widetilde{x}|} \sin {\varphi} \left( \, { \frac{\widetilde{t}-\widetilde{x}}{1-\widetilde{x}}} \, \right)   \, ,
\label{eq:stft8}
\end{equation}
which is the exact same result obtained in Eq.~(\ref{eq:final2}).

\section{Analytical vs Numerical Solution}
\label{sec:solutions}

For the physically common dispersion relation $\widetilde{\omega} = {\widetilde{k}}^2$, the dimensionless angular frequencies at the source can be obtained using Eq.~(\ref{eq:velocity2}) as,
\begin{equation}
\widetilde{\omega}(\widetilde{t}_s)  = \left ( \frac{k_0 L}{\omega_0 t_0} \frac{1}{1-\widetilde{t}_s} \right)^2 \, ,
\label{eq:frequencies}
\end{equation}
which corresponds to the angular phase
\begin{equation}
{\varphi}(\widetilde{t}_s)  =  \frac{k_0^2 L^2}{\omega_0 t_0} \frac{\widetilde{t}_s}{1-\widetilde{t}_s} \, .
\label{eq:phases}
\end{equation}
where the integration constant is fixed in such a way that $\varphi(0) = 0$. This result can be extended to a more general dispersion relation, $\widetilde{\omega} = \widetilde{k}^\alpha$ ($\alpha \neq 1$), in which the angular phase becomes:
\begin{equation}
    \varphi(\widetilde{t}_s)  =  \omega_0 t_0 (\alpha -1) \left(\frac{k_0 L}{\omega_0 t_0} \right)^{\frac{\alpha}{\alpha -1}} \left[ (1- \widetilde{t}_s)^{\frac{1}{1-\alpha}} - 1 \right] \, .
\label{eq:generalangularphase}    
\end{equation}
It must be stressed that the method fails for $\alpha =1$, which corresponds to the case of a non-dispersive medium.

\begin{figure}[h]
    \centering
\includegraphics[width=0.55\textwidth]{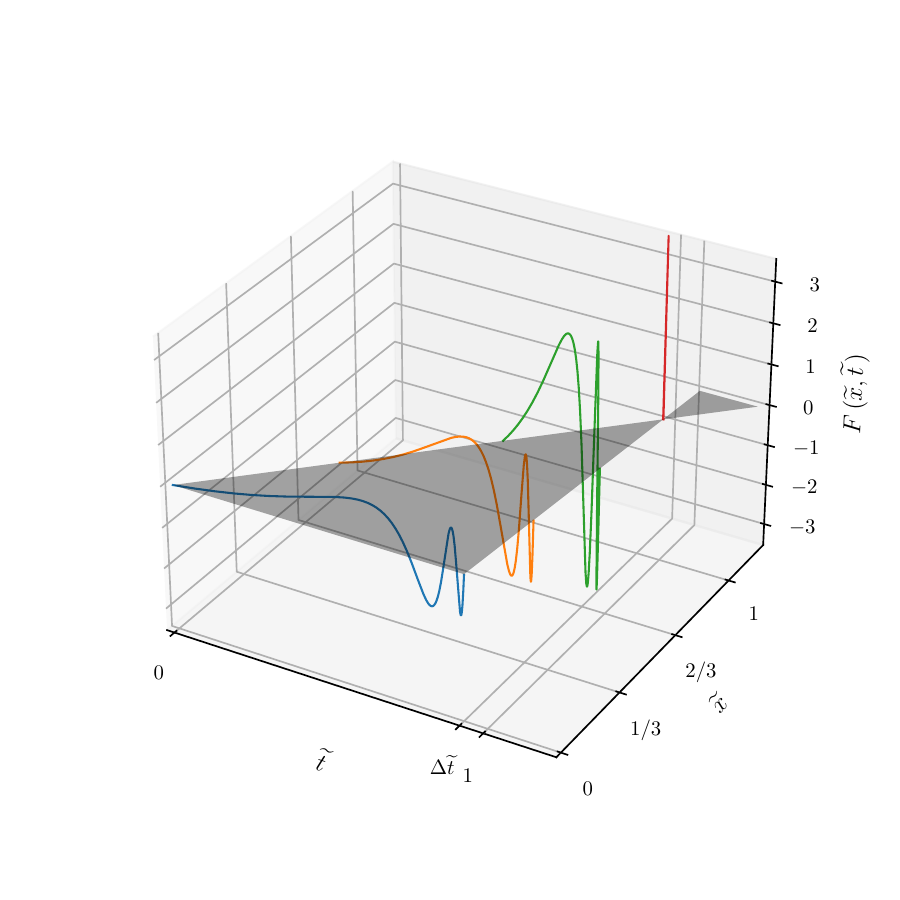}
    \caption{Time profile of $F(\, \widetilde{x},\widetilde{t}\, )$ for the dispersion relation $\widetilde{\omega} = \widetilde{k}^2$, as it evolves through the spatial coordinate, compressing and increasing its amplitude, from the source at $\widetilde{x}=0$ (blue) through $\widetilde{x}=1/3$ (orange) and $\widetilde{x}=2/3$ (green), until the diverging point at $\widetilde{x}=1$ (red), \emph{i.e.} the Dirac-delta function. The value of $\Delta \widetilde{t}$ is set at approximately 0.9263.}
    \label{fig:diagram3D}
\end{figure}

Assuming once more the special case where the values of $\omega_0 t_0$ and $k_0 L$ are equal to one, the analytical solution can be presented in a compact form as,
\begin{equation}
\widetilde{\omega}= \widetilde{k}^2   \Rightarrow F(\,\widetilde{x},\widetilde{t}\,)=\frac{1}{|1-\widetilde{x}|} \sin\left( \, \frac{\widetilde{t}-\widetilde{x}}{1-\widetilde{t}} \, \right) \, .
\label{eq:analyticalsquared}
\end{equation}
This result can be visualized in Figure \ref{fig:diagram3D}, which shows the time profile of the wave as it evolves through the spatial coordinate, compressing and increasing its amplitude. The value of $\Delta \widetilde{t}$ is set at approximately 0.9263 for visual effect purposes, \emph{i.e.} so the signal includes two full cycles, starting and ending at zero.

For any point $\widetilde{x}$ between $0$ and $1$, it can be seen that the time dependence of the amplitude has the exact same form as the time profile at $\widetilde{x}=0$, within the allowed time interval. After all, this is the only possible way that such point could operate as a source and lead to the same localized wave at $\widetilde{x}=1$. In other words, the amplitude at the source is being mapped to any position $\widetilde{x}$ with the following transform:
\begin{equation}
\widetilde{t}_s \rightarrow \frac{\widetilde{t}-\widetilde{x}}{1-\widetilde{x}}\, ,
\end{equation}
shifting the function to the right by an amount $\widetilde{x}$ and compressing it by a factor $1-\widetilde{x}$. As the integral remains constant due to the superposition principle, the amplitude also scales with the same factor.

At this point, it is convenient to revisit the progression of the wave in space as shown in Fig.~\ref{fig:diagram3D}. For the sake of clarity, one can assume the wave in the example is mechanical and the amplitude corresponds to a physical displacement. At the source, \emph{i.e.} at the fixed location $\widetilde{x}=0$, an object riding the wave will complete two cycles in the dimensionless time interval $\Delta\widetilde{t}$. However, at any other fixed location, for example, $\widetilde{x}=1/3$ or $\widetilde{x}=2/3$, an object riding the wave will complete the very same cycles in a shorter and shorter time, respectively. In the extreme case of $\widetilde{x}=1$, the cycles would be completed instantaneously. This leads to an apparent contradiction. When a sinusoidal wave with a fixed frequency travels into a dispersive medium, the frequency must be preserved while the wavenumber changes according to the dispersion relation. Nonetheless, in this example, the frequency appears to increase as the wave progresses into the medium. This is not the case, however, as each component of the STFT maintains its own frequency while traveling through the dispersive medium. The apparent increase in the frequency of the object riding the wave is no more than a collective effect of the superposition of these components.

\subsection{Numerical solution}

Despite the fact that this analytical solution is supported by  two different formal derivations, it is critical to test this result with an independent brute force numerical algorithm for confirmation. The procedure for the numerical approach is presented below and compared with the analytical solution of the physical case considered before.

\begin{figure}[h]
    \centering
\includegraphics[width=0.55\textwidth]{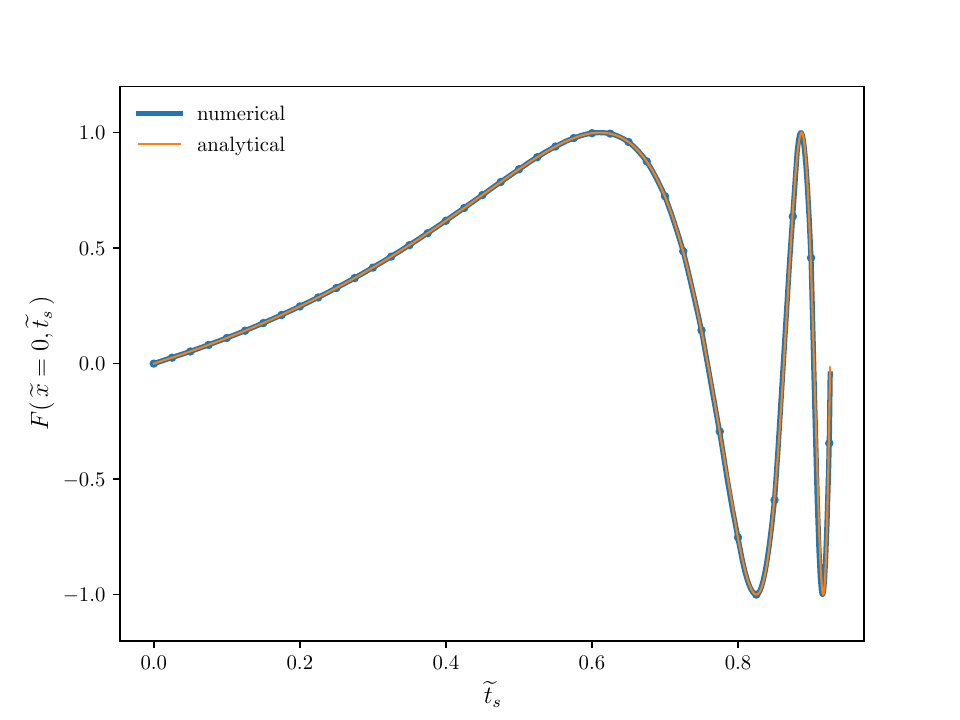}
    \caption{Comparison between the analytical function at the source, $F(\, \widetilde{x}=0,\widetilde{t}_s\, )$, represented by the orange line, with the discrete sum of sinusoidal segments shown in blue. The blue dots represent the beginning and end of each segment for $\Delta\widetilde{t}_s=0.025$.}
    \label{fig:sourcenumerical}
\end{figure}

The mathematical function described by Eq.~(\ref{eq:sourcewave}) represents the source of a sinusoidal plane wave with a varying frequency. In the discrete limit, this can be approximated as a series of sinusoidal segments, where each segment has a constant frequency,
\begin{equation}
F_n(\, \widetilde{x}=0,\widetilde{t}_s\, ) = \sin\left[ \widetilde{\omega}(\widetilde{t}_n) \widetilde{t}_s + \phi_n \right] \, , \,\,\,\,\,\,\,\,\,\,\,\,\,\,\,\, \widetilde{t}_n - \frac{\Delta \widetilde{t}_s}{2} < \widetilde{t}_s < \widetilde{t}_n + \frac{\Delta \widetilde{t}_s}{2} \, ,
\end{equation}
where $\Delta \widetilde{t}_s$ is the dimensionless time interval of each segment and,
\begin{equation}
\widetilde{t}_n = \left( n - \frac{1}{2} \right) \Delta \widetilde{t}_s \, , \,\,\,\,\,\,\,\,\,\,\,\,\,\,\,\, n = 1 \,, \, 2\,, \, ...\, , N 
\end{equation}
corresponds to the time instant at the center of the $n$th segment, with $N=\Delta \widetilde{t}/\Delta \widetilde{t}_s$, where $\Delta \widetilde{t}_s$ is chosen in such a way that $N$ is an integer. The dimensionless angular frequency is determined by Eq.~(\ref{eq:velocity2}) and by the dispersion relation specific to medium. 
These relations ensure that each segment arrives at $\widetilde{x}=1$ simultaneously at $\widetilde{t}=1$. In order to maintain continuity, the phase of each sinusoidal segment can be determined recursively as,
\begin{equation}
\widetilde{\omega}(\widetilde{t}_n) n\Delta\widetilde{t}_s + \phi_n = \widetilde{\omega}(\widetilde{t}_{n+1}) n \Delta\widetilde{t}_s + \phi_{n+1} \, .
\end{equation}
As an example, the agreement between the analytical continuum limit of the varying frequency wave at the source with the discrete sum of sinusoidal segments can be observed in Fig.~\ref{fig:sourcenumerical} for the dispersion relation $\widetilde{\omega} = \widetilde{k}^2$ and $\Delta\widetilde{t}_s=0.025$. The blue dots represent the beginning and end of each segment.

\begin{figure}[!h]
\centering
\includegraphics[width=0.325\textwidth]{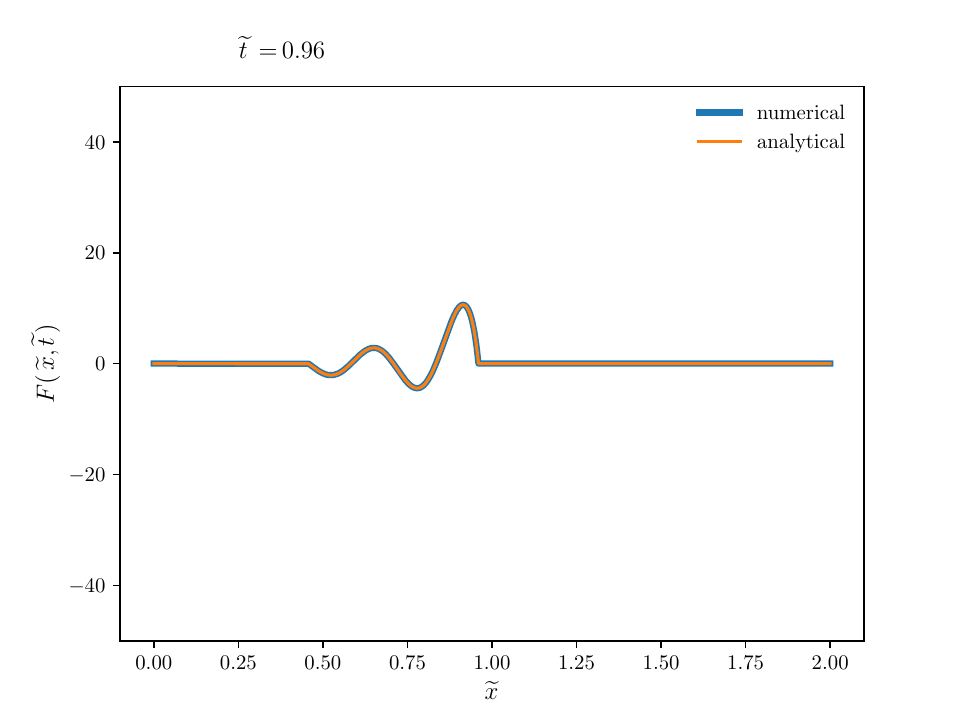}
\includegraphics[width=0.325\textwidth]{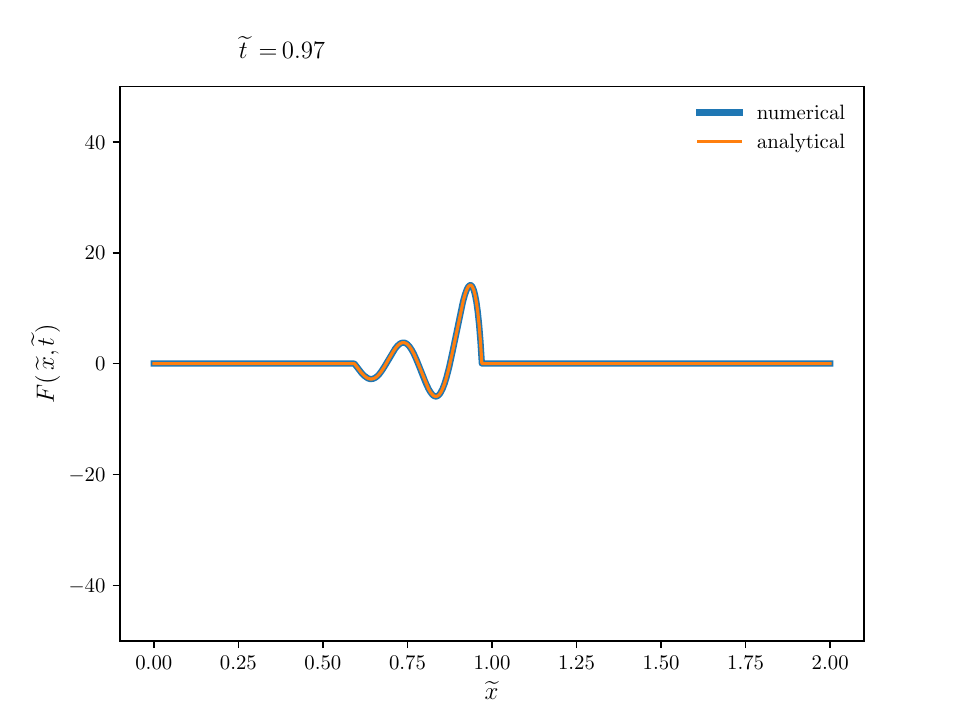}
\includegraphics[width=0.325\textwidth]{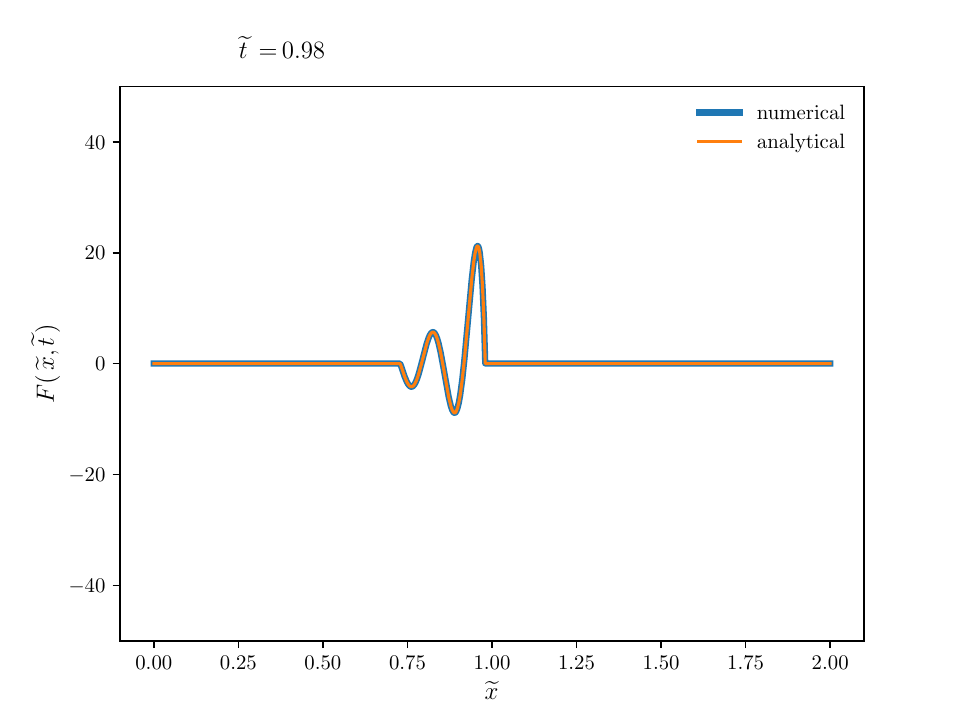} \\
\includegraphics[width=0.325\textwidth]{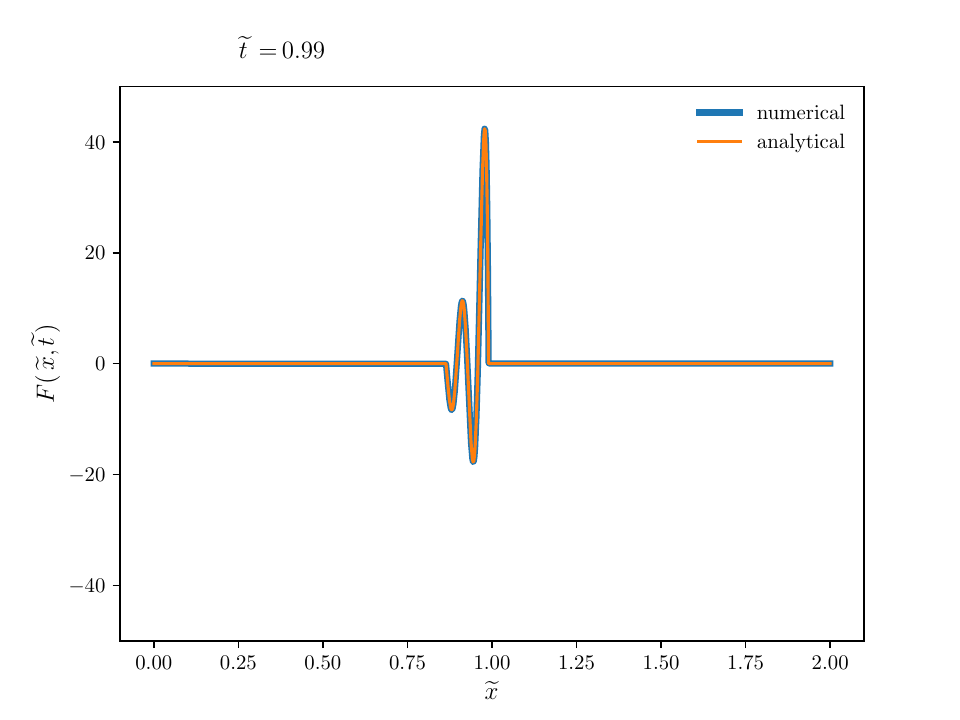}
\includegraphics[width=0.325\textwidth]{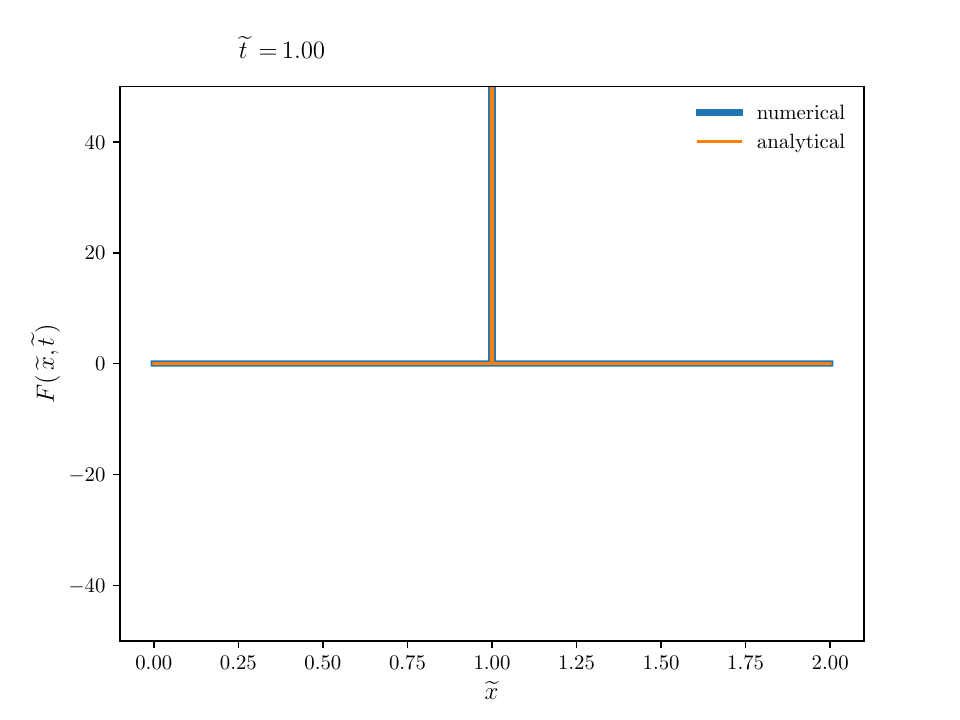}
\includegraphics[width=0.325\textwidth]{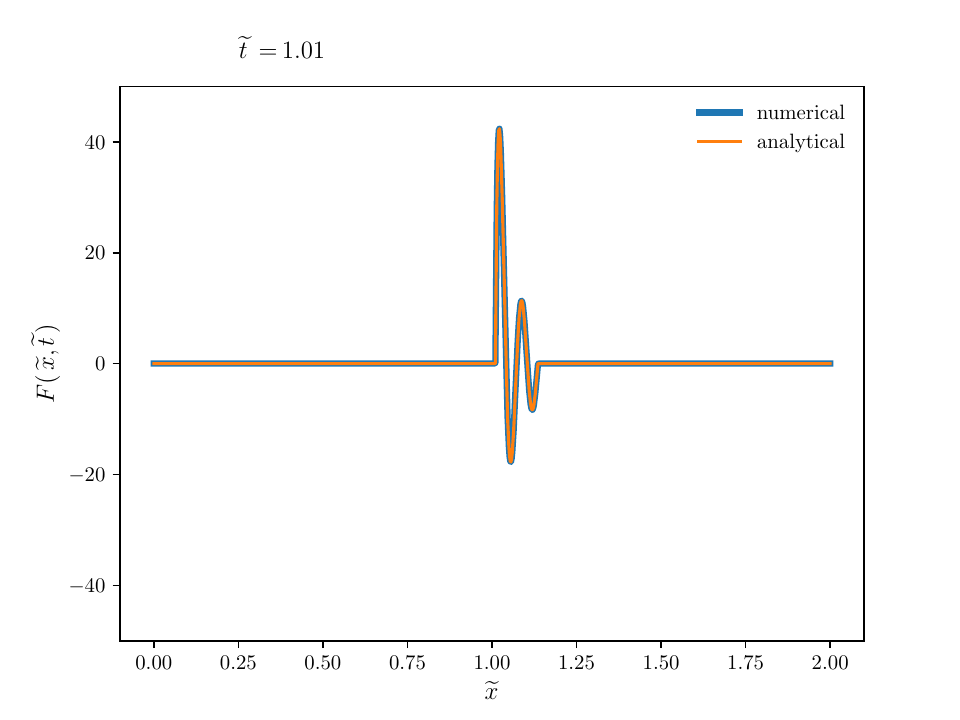} \\
\includegraphics[width=0.325\textwidth]{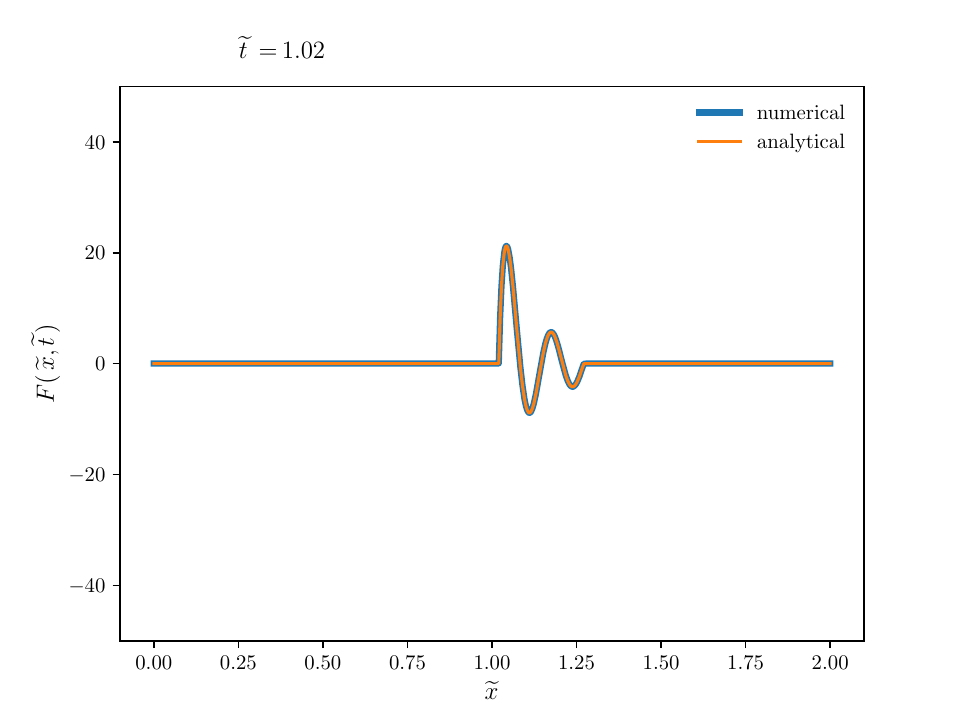}
\includegraphics[width=0.325\textwidth]{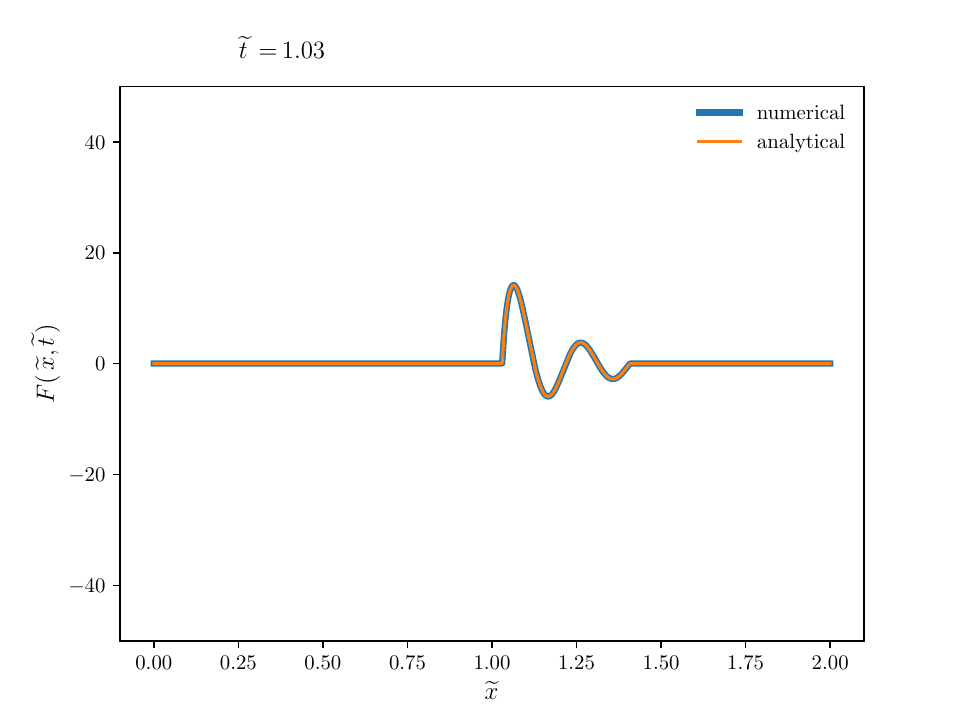}
\includegraphics[width=0.325\textwidth]{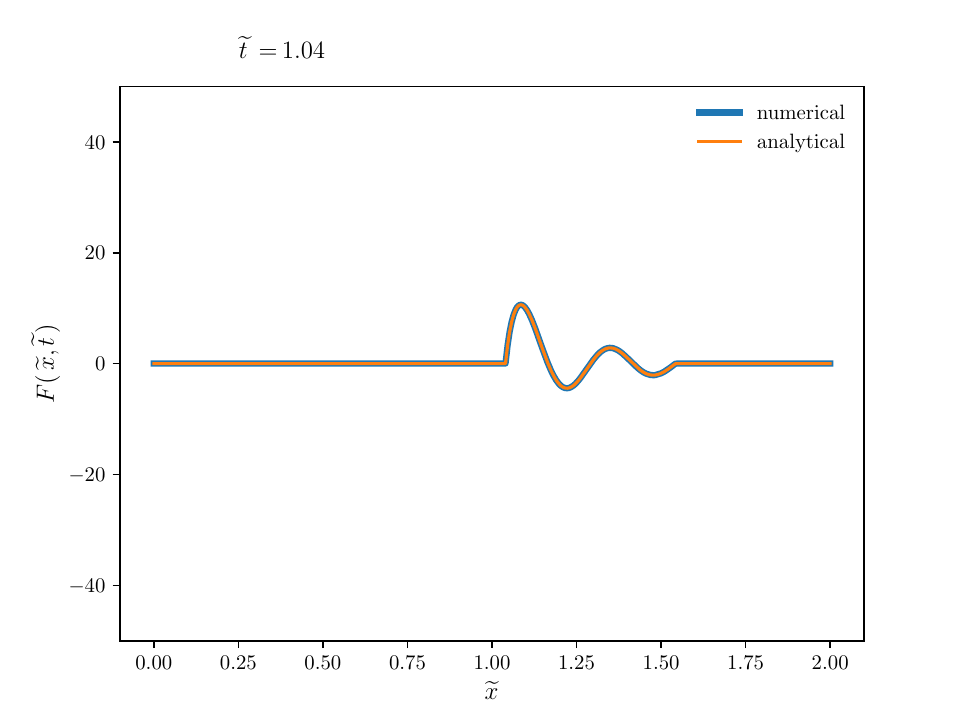}
\caption{Comparison between the numerical and the analytical solutions for dimensionless time instants ranging from 0.96 to 1.04. The figure refers to the case of the dispersion relation $\widetilde{\omega} = \widetilde{k}^2$.}
\label{fig:comparison}
\end{figure}

As each time interval corresponds to a sinusoidal plane wave with a fixed frequency being radiated from the source, the overall wave in the numerical approximation can be obtained as the superposition of each segment moving through space with their corresponding constant velocity,
\begin{equation}
F(\,\widetilde{x},\widetilde{t}\,)=  \sum_{n=1}^N F_n(\,\widetilde{x},\widetilde{t}\,) = \sum_{n=1}^N \mathcal{U}_n \sin\left[ \widetilde{\omega}(\widetilde{t}_n) \widetilde{t} - \widetilde{k}(\widetilde{t}_n)  \widetilde{x} +\phi_n \right] \, ,
\end{equation}
where,
\begin{eqnarray}
\mathcal{U}_n=
    \begin{cases}
        1 & \text{if } \widetilde{x} \left [ 1- \left( n - \frac{1}{2} \right) \Delta\widetilde{t}_s \right] + \left( n-1 \right) \Delta\widetilde{t}_s < \widetilde{t} < \widetilde{x} \left [ 1- \left( n - \frac{1}{2} \right) \Delta\widetilde{t}_s \right] + n \Delta\widetilde{t}_s \, , \\
        0 & \text{otherwise.}
    \end{cases}
\end{eqnarray}
It should be noted that in this numerical approach, as the signal is divided into shorter segments of equal length, and approximated as a sinusoidal function in each segment, $\mathcal{U}_n$ can be interpreted as a rectangular window function. As the segments progress in space and time as plane waves, their overlap will naturally lead to discontinuities at the edges of the segments as the overall wave starts compressing. In order to address this issue, a Constant Overlap-Add (COLA) window function, such as the Hann function~\cite{smith3}, is employed in the numerical calculation leading to Fig.~\ref{fig:comparison}, in order to ensure the overall wave's smoothness. It is worth stressing that as the time interval of the segments approaches zero, the choice of the window function becomes inconsequential.

The comparison between the numerical and the analytical solution is presented in Fig.~\ref{fig:comparison} for $\Delta\widetilde{t}_s=10^{-4}$. The comparison is shown for different time instants right before and after the formation of the Dirac-delta peak, presenting a nearly perfect agreement. The dimensionless time instants in the panels range from 0.96 to 1.04, corresponding to the interval where the wave compresses above tenfold the initial amplitude and diverges at 1.00. The complete time progression of the wave can also be seen in the following video~\cite{video}, again showing the excellent agreement between the analytical and numerical results. In both representations, it is visible that the wave presents symmetric shapes before and after $\widetilde{t}=1$. For example, the shape at $\widetilde{t}=0.96$ is mirrored with respect to $\widetilde{x}=1$ when compared with the shape at $\widetilde{t}=1.04$.

\section{Discussion and Conclusions}
\label{sec:discussion}

Even though the previous numerical analysis focused on the case of a particular dispersion relation, the same approach can be applied to any physical scenario where dispersion occurs. In fact, the angular phase required to produce a localized Dirac-delta wave was obtained for a general power function dispersion relation in Eq.~(\ref{eq:generalangularphase}). This is the case, for example, of water waves in the deep water regime where the dispersion relation is approximately $\widetilde{\omega}= {\widetilde{k}}^{1/2}$, according to the Airy wave theory. As the frequency of water waves depends primarily on the wind speed, it is plausible that a varying gust in the right conditions may lead to the formation of a rogue wave in a similar fashion as the localized Dirac-delta wave.

Creating a focused wave targeting a specific region of space holds immense potential in a number of applications. In the realm of cancer treatment for example, the ability to direct waves precisely to the affected area allows for more effective and less invasive therapies. As several types of waves experience dispersion in biological tissues~\cite{schwan1,schwan2,guedert,sugimoto}, a targeted approach like the one proposed in this paper could potentially minimize collateral damage to surrounding healthy tissues in the medical treatment of tumors, enhancing the precision of medical interventions. Another appealing physical scenario is the case of ion-acoustic waves in  fusion plasmas. Ion acoustic waves are a distinctive form of plasma oscillations that propagate through ionized gases, or plasmas, found in a variety of astrophysical and laboratory settings~\cite{gekelman}. These waves arise from the collective motion of ions in response to perturbations, generating compression disturbances in the plasma. Dominated by the inertia of ions and the restoring force due to thermal pressure, ion acoustic waves play a crucial role in fusion experiments. While the dispersion of ion-acoustic waves is more complex than the case analyzed in this paper~\cite{fitzpatrick,liu}, the linear nature of such waves allows for the possibility of producing a localized Dirac-delta wave with an appropriate choice for the time-dependent angular phase at the source. The production of such wave could possibly lead to a localized increase of density above the threshold for nuclear fusion reactions to occur, enabling a controlled release of energy, leaving the rest of the plasma mostly unaffected.

\begin{figure}[h]
    \centering
\includegraphics[width=0.55\textwidth]{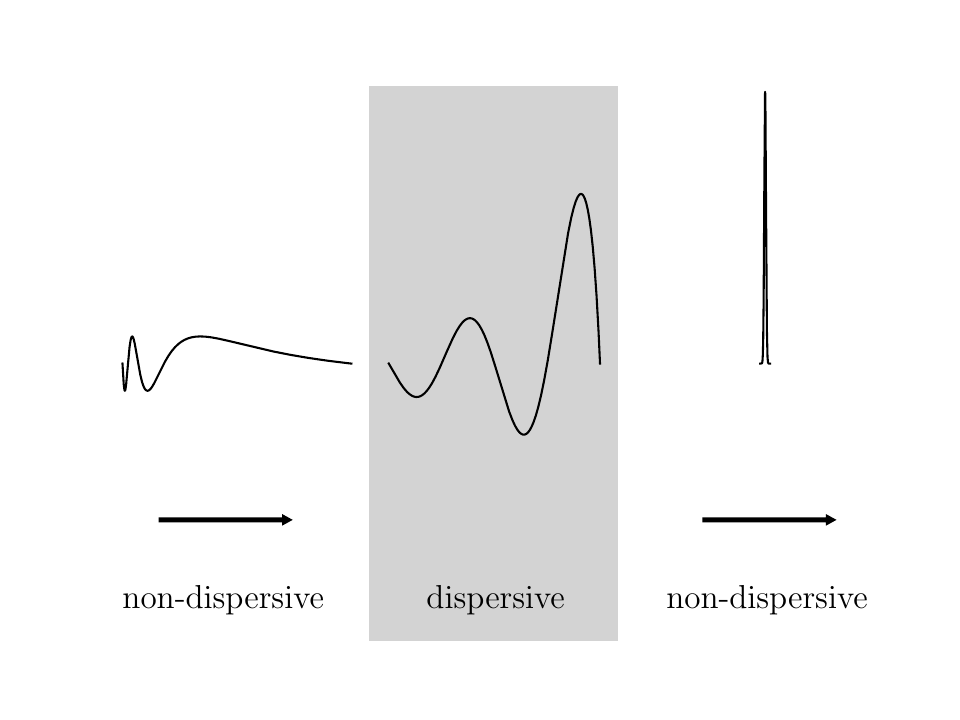}
    \caption{Progression of a chirped sinusoidal wave from a non-dispersive medium into a dispersive slab. The thickness is such that the wave compresses into a Dirac-delta at the end of the slab, and exits back into the non-dispersive medium as a Dirac-delta pulse wave.}
    \label{fig:diagramslab}
\end{figure}

A further compelling consequence of localized Dirac-delta waves is the ability to produce an impulse signal. It is widely known that a chirp shares the same power spectrum of an impulse signal~\cite{smith}. Therefore, when an impulse signal is transmitted into a dispersive medium, it may undergo a conversion into a chirp. This, in fact, is the principle behind Chirped Pulse Amplification (CPA)~\cite{strickland}, where an ultra-short laser pulse is temporally extended (chirped) to reduce its peak power using diffraction gratings. Subsequently, it is amplified without causing damage to the laser medium. Following amplification, the pulse is compressed back to its original duration, resulting in an intense ultrashort laser pulse. This technique ultimately led to the Nobel Prize in Physics in 2018, awarded to Strickland and Mourou. Analogously, a chirp can also be propagated into a dispersive medium in a manner that momentarily converges the signal into a Dirac-delta function before dispersing into the background, much like a rogue wave. As such, if a chirp signal is sent from a non-dispersive medium into a dispersive slab with the exact thickness to converge into a Dirac-delta at the end of the slab, the signal would be transmitted as a Dirac-delta pulse back into the non-dispersive medium, as exemplified in Fig.~\ref{fig:diagramslab}. Therefore, the production of the localized Dirac-delta wave as detailed in this paper can also be interpreted as the time reversal ($t\rightarrow -t$) of a Dirac-delta pulse emitted into a dispersive medium. As a result, it shall be possible to use this time symmetry to find out experimentally which shape a signal should have at the source in order to produce a Delta-function pulse at the desirable distance from the source. This may be particularly useful in media with complicated dispersion relations where the analytical calculation is challenging.

In summary, this study presents a novel solution to the linear wave equation in dispersive media: a localized Dirac-delta wave. The proposed solution arises from the interference of a sinusoidal wave characterized by a time-dependent frequency, uniquely dictated by the dispersion relation of the medium. Both numerical and exact analytical solutions are presented for a particular physical scenario, showing a nearly perfect agreement between them. While a thorough exploration of the practical applications stemming from this result is not within the scope of this paper, the authors are optimistic that the proposal may inspire and contribute to potential real-world implementations.




\section*{Acknowledgments}

The authors would like to thank Dr.~Giovanni Ossola for carefully reading the manuscript and giving valuable suggestions.


\section*{Appendix: Dirac Delta function properties of the function $F$}

In this subsection it is shown that the integral of $F$ at fixed $\widetilde{t}$ does not depend on $\widetilde{t}$, as long as {$ \Delta \widetilde{t} < \widetilde{t} < 1$}. The integral of interest is  
\begin{equation}
    G_x = \int_{\frac{\widetilde{t} - \Delta \widetilde{t}}{1 -\Delta \widetilde{t}}}^{\widetilde{t}}  \, \frac{1}{1-\widetilde{x}} \sin{ \varphi \left(\frac{\widetilde{t} - \widetilde{x}}{1 - \widetilde{x}} \right) } \, \textrm{d} \widetilde{x}\, .
\end{equation}
It is now sufficient to change integration variable according to
\begin{equation}
\widetilde{t}_s = \frac{\widetilde{t} - \widetilde{x}}{1 - \widetilde{x}} \, , \qquad \widetilde{x} = \frac{\widetilde{t} - \widetilde{t}_s}{1 - \widetilde{t}_s} \, , \qquad  \textrm{d} \widetilde{x} = - \frac{1-\widetilde{t}}{(1- \widetilde{t}_s)^2} \textrm{d} \widetilde{t}_s \, ,
\end{equation}
to obtain
\begin{equation}
    G_x(\, \Delta \widetilde{t} \, ) = \int_0 ^{\Delta \widetilde{t}} \frac{1}{1-\widetilde{t}_s}  \sin{\varphi(\widetilde{t}_s) } \, \textrm{d} \widetilde{t}_s \, ,
\end{equation}
where it was made explicit that $G$ is a function of the parameter $\Delta \widetilde{t}$. It is therefore obvious that 
\begin{equation}
\frac{1}{G_x( \,\Delta \widetilde{t} \,)}\int_{\frac{\widetilde{t} - \Delta \widetilde{t}}{1 -\Delta \widetilde{t}}}^{\widetilde{t}}  \frac{1}{1-\widetilde{x}} \sin{ \varphi \left(\frac{\widetilde{t} - \widetilde{x}}{1 - \widetilde{x}} \right) }\, \textrm{d} \widetilde{x} =  1 \,.
\end{equation}

With the same change of variables discussed above, it is possible to prove that for a sufficiently smooth function $h(\widetilde{x})$
\begin{equation}
 I =    \lim_{\widetilde{t} \to 1} \frac{1}{G_x( \,\Delta \widetilde{t} \,)} \int_{\frac{\widetilde{t} - \Delta \widetilde{t}}{1 -\Delta \widetilde{t}}}^{\widetilde{t}} \, h(\widetilde{x})\frac{1}{1-\widetilde{x}} \sin{ \varphi \left(\frac{\widetilde{t} - \widetilde{x}}{1 - \widetilde{x}} \right) } \, \textrm{d} \widetilde{x} = h(1) \, .
\end{equation}
Indeed, after the change in integration variables $\widetilde{x} \to \widetilde{t}_s$ introduced above, one finds 
\begin{equation}
    I =  \lim_{\widetilde{t} \to 1} \frac{1}{G_x( \,\Delta \widetilde{t} \,)} \int_0 ^{\Delta \widetilde{t}}  h \left( \frac{\widetilde{t} - \widetilde{t}_s}{1 -\widetilde{t}_s}\right)\frac{1}{1-\widetilde{t}_s}  \sin{\varphi(\widetilde   {t}_s) } \, \textrm{d} \widetilde{t}_s \, .
\end{equation}
At this stage, it is possible to take the limit of the integrand before integration, to find 
\begin{equation}
    I = \frac{1}{G_x( \,\Delta \widetilde{t} \,)} \int_0 ^{\Delta \widetilde{t}}  \, h \left( 1\right)\frac{1}{1-\widetilde{t}_s}  \sin{\left(\varphi(\widetilde{t}_s) \right)} \, \textrm{d} \widetilde{t}_s  = h (1).
\end{equation}
It is therefore possible to conclude that 
\begin{equation}
\lim_{\widetilde{t} \to 1} \frac{1}{G_x( \,\Delta \widetilde{t} \,)}  \int_{\frac{\widetilde{t} - \Delta \widetilde{t}}{1 -\Delta \widetilde{t}}}^{\widetilde{t}} h(\widetilde{x}) \frac{1}{1-\widetilde{x}} \sin{ \varphi \left(\frac{\widetilde{t} - \widetilde{x}}{1 - \widetilde{x}} \right) } \, \textrm{d} \widetilde{x}   =  \int_{-\infty}^{\infty}  h(\widetilde{x}) \delta(1-\widetilde{x}) \, \textrm{d} \widetilde{x} = h(1) \, .
\end{equation}


Similar considerations apply to the integrals of the function $F$ with respect to $\widetilde{t}$. For instance, one can also show that the integral of $F$ at fixed $\widetilde{x}$ does not depend on $\widetilde{t}$. In this case, the integral of interest is  
\begin{equation}
    G_t = \int_{\widetilde{x}}^{\widetilde{x}+\Delta\widetilde{t} (1-\widetilde{x})} \frac{1}{1-\widetilde{x}} \sin{ \varphi \left(\frac{\widetilde{t} - \widetilde{x}}{1 - \widetilde{x}} \right) } \,  \textrm{d} \widetilde{t}  \, .
\end{equation}
It is now sufficient to change integration variable according to
\begin{equation}
\widetilde{t}_s = \frac{\widetilde{t} - \widetilde{x}}{1 - \widetilde{x}} \, , \qquad \widetilde{t} = \widetilde{x} + \widetilde{t}_s (1-\widetilde{x} )  \, , \qquad  \textrm{d} \widetilde{t} = (1-\widetilde{x})  \textrm{d} \widetilde{t}_s \, ,
\end{equation}
to obtain
\begin{equation}
    G_t(\, \Delta \widetilde{t} \, ) = \int_0 ^{\Delta \widetilde{t} }  \sin{\varphi(\widetilde{t}_s) } \, \textrm{d} \widetilde{t}_s  \, ,
\end{equation}
where it was made explicit that $G$ is a function of the parameter $\Delta \widetilde{t} $. It is therefore obvious that 
\begin{equation}
\frac{1}{G_t( \,\Delta \widetilde{t} \,)}\int_{\widetilde{x}}^{\widetilde{x}+\Delta\widetilde{t} (1-\widetilde{x})} \frac{1}{1-\widetilde{x}} \sin{ \varphi \left(\frac{\widetilde{t} - \widetilde{x}}{1 - \widetilde{x}} \right) } \, \textrm{d} \widetilde{t} =  1 \,.
\end{equation}

With the same change of variables discussed above, it is possible to prove that for a sufficiently smooth function $h(\widetilde{x})$
\begin{equation}
 I =    \lim_{\widetilde{x} \to 1} \frac{1}{G_t( \,\Delta \widetilde{t} \,)} \int_{\widetilde{x}}^{\widetilde{x}+\Delta\widetilde{t} (1-\widetilde{x})}  h(\,\widetilde{t}\,)\frac{1}{1-\widetilde{x}} \sin{ \varphi \left(\frac{\widetilde{t} - \widetilde{x}}{1 - \widetilde{x}} \right) } \, \textrm{d} \widetilde{t}  = h(1) \, .
\end{equation}
Indeed, after the change in integration variables $\widetilde{x} \to \widetilde{t}_s$ introduced above, one finds 
\begin{equation}
    I =  \lim_{\widetilde{x} \to 1} \frac{1}{G_t( \,\Delta \widetilde{t} \,)} \int_0 ^{\Delta \widetilde{t}}  h \left( \widetilde{x} + \widetilde{t}_s (1-\widetilde{x} ) \right)  \sin{\varphi(\widetilde   {t}_s) } \,  \textrm{d} \widetilde{t}_s \, .
\end{equation}
It is again  possible to take the limit of the integrand before integration, to find 
\begin{equation}
    I = \frac{1}{G_t( \,\Delta \widetilde{t} \,)} \int_0 ^{\Delta \widetilde{t}} h \left( 1\right)  \sin{\varphi(\widetilde{t}_s) } \, \textrm{d} \widetilde{t}_s = h (1).
\end{equation}
Consequently, one finds that 
\begin{equation}
\lim_{\widetilde{x} \to 1} \frac{1}{G_t( \,\Delta \widetilde{t} \,)}  \int_{\widetilde{x}}^{\widetilde{x}+\Delta\widetilde{t} (1-\widetilde{x})}  h(\,\widetilde{t}\,) \frac{1}{1-\widetilde{x}} \sin{ \varphi \left(\frac{\widetilde{t} - \widetilde{x}}{1 - \widetilde{x}} \right) } \, \textrm{d} \widetilde{t}   =  \int_{-\infty}^{\infty}h(\,\widetilde{t}\,) \delta(\,1-\widetilde{t}\,) \, \textrm{d} \widetilde{t} = h(1) \, .
\end{equation}



\end{document}